\documentclass[twocolumn,showpacs,amsmath,amssymb,aps,prd,floats,nofootinbib]{revtex4-2}
\usepackage{graphicx}
\usepackage[switch,pagewise,running]{lineno}
\usepackage{dcolumn}
\usepackage{bm}
\usepackage{xspace}
\usepackage{hyperref}
\hypersetup{colorlinks,linkcolor=blue,citecolor=blue,urlcolor=blue}
\usepackage{color,xcolor}
\usepackage{booktabs}
\usepackage{placeins}

\renewcommand{\arraystretch}{1.25}

\begin{document}

\title{A Model for Magnetic Reconnection as the Origin of TeV Outbursts from NGC~1275}

\author{Zhen-Jie Wang$^{1,2}$, Hong-Bin Tan$^{1,2}$, Ruo-Yu Liu$^{1,2,3}$}
\affiliation{$^1$School of Astronomy and Space Science, Nanjing University, 210023 Nanjing, Jiangsu, China; \textcolor{blue}{ryliu@nju.edu.cn}\\
$^2$Key Laboratory of Modern Astronomy and Astrophysics (Nanjing University), Ministry of Education, Nanjing 210023, China\\
$^3$Tianfu Cosmic Ray Research Center, Chengdu 610000, Sichuan, China\\}

\begin{abstract}
NGC~1275 showed two TeV $\gamma$-ray outbursts between November 2022 and January 2023, as detected by the Large High Altitude Air Shower Observatory (LHAASO).  The source was also active in the X-ray and GeV bands during the TeV outburst period.  Very-long-baseline radio observations reported a sudden acceleration and deflection of a jet knot in late 2022, before the main TeV activity.  Motivated by this sequence, we examine whether magnetic reconnection triggered by the interaction between the jet and the ambient medium can explain the TeV flares.  In this picture, reconnection produces many plasmoids, and a large ``monster'' plasmoid becomes the main flare region.  We model the low-state emission with a multi-zone stochastic-dissipation component and add a compact reconnection-powered region for the flaring state.  We then compare leptonic and hadronic interpretations.  The leptonic model explains the enhanced X-ray emission as electron synchrotron radiation and the TeV emission mainly as inverse-Compton radiation.  A pure proton--proton model can also produce TeV photons if dense target gas is present, but it requires a compact cloud with a density above the values directly inferred from free--free absorption and a large proton power.  These requirements are demanding, but they do not by themselves exclude the hadronic interpretation, because the gas may be compressed by the jet or may contain denser cloud cores.  Our results show that magnetic reconnection in the parsec-scale jet is a viable origin of the 2022--2023 TeV activity of NGC~1275, while better constraints on the gas density and jet power are needed to distinguish between leptonic and hadronic radiation channels.
\end{abstract}

\maketitle

\section{Introduction}

Relativistic jets from active galactic nuclei (AGNs) are powerful non-thermal sources.  If a jet points close to our line of sight, Doppler boosting makes the source appear as a blazar.  If a similar jet is viewed at a larger angle, the source can appear as a radio galaxy in the standard unification picture \citep{1995PASP..107..803U, Blandford_2019ARA&A..57..467B}.  Blazars and misaligned radio galaxies are therefore useful for studying the same basic jet physics under different viewing conditions.

The broadband spectral energy distributions (SEDs) of these sources usually have two broad components.  The low-energy component is produced by synchrotron radiation from relativistic electrons.  The origin of the high-energy component, especially in the GeV--TeV band, is less certain.  In leptonic models, the high-energy photons are produced by inverse-Compton (IC) scattering of synchrotron or external photons.  In hadronic models, the emission may come from proton synchrotron radiation, photohadronic interactions, or proton--proton (pp) collisions, with possible electromagnetic cascades and neutrino production \citep{Konopelko_2003ApJ...597..851K, Bottcher_2013, Wang_2024ApJS..271...10W, Dutta_2024ApJ...974...56D, Xue_2022PhRvD.106j3021X}.  To distinguish these possibilities, one needs constraints on the variability time, Doppler factor, magnetic field, target photon or gas density, and jet power.

Fast TeV variability also constrains how the jet dissipates energy.  Magnetic reconnection is a natural mechanism for converting magnetic energy into particle energy in compact regions of relativistic jets \citep{Giannios_2009MNRAS, Giannios_2013MNRAS.431..355G, Uzdensky_2010PhRvL.105w5002U, Sironi_2025ARA&A..63..127S}.  Reconnection can form a chain of plasmoids.  A small number of large plasmoids may dominate short flares because they contain more particles and magnetic energy, and because their local Doppler factors can be enhanced.  This idea has been used to explain rapid high-energy flares in several blazars \citep{Aharonian_2007, Aleksi_2011ApJ...730L...8A, Jorstad_2013ApJ...773..147J}.

NGC~1275 (Perseus~A, 3C~84) is a nearby radio-loud AGN at redshift $z=0.0176$ \citep{Strauss_1992ApJS...83...29S}.  It is associated with an FR~I radio galaxy/Seyfert-like nucleus \citep{Chiaberge_1999A&A...349...77C} and hosts a central black hole of mass $M_{\rm BH}=(0.8-2)\times10^{9}M_{\odot}$ \citep{Scharwachter_2013MNRAS.429.2315S, Giovannini_2018NatAs...2..472G}.  Although the jet is not as closely aligned with the line of sight as in classical blazars, NGC~1275 is detected from radio to TeV energies.  It is therefore a nearby system in which blazar-like jet physics can be tested in a misaligned geometry.  The source was first detected in high-energy $\gamma$ rays by \emph{Fermi}-LAT \citep{Abdo_2009ApJ_699_31}, and very-high-energy emission above $100\,$GeV was later reported by MAGIC \citep{Aleksic_2012AA_539_L2}.  Multi-wavelength campaigns have shown that the GeV--TeV emission is variable and is related to compact jet activity \citep{Aleksic_2014AA_564_A5, Dutson_2014MNRAS_442_2048}.

Between November 2022 and January 2023, the Water Cherenkov Detector Array (WCDA) of LHAASO detected two TeV $\gamma$-ray outbursts from NGC~1275 \citep{Caozhen_2025MNRAS.540.1860C}. The first outburst lasted from Modified Julian Date (MJD) 59933.445 to 59935.439 (hereafter, O1) and the second one lasted from MJD 59951.396 to 59958.376 (hereafter, O2). The source also brightened in the GeV $\gamma$-ray and X-ray bands during this period \citep{Caozhen_2025MNRAS.540.1860C, Godambe_2024ApJ...974L..31G}.  Radio observations reported a sudden acceleration and a change in direction of a jet knot in late 2022, before the main TeV outbursts \citep{Park_2024A&A...685A.115P}.  This timing does not prove a causal connection, but it motivates us to think about that can a rapid change in the parsec-scale jet trigger magnetic reconnection and produce the TeV flares?

These TeV flares have already been discussed in simple radiative models.  The LHAASO discovery paper showed that a one-zone synchrotron self-Compton (SSC) model can reproduce the $\gamma$-ray data \citep{Caozhen_2025MNRAS.540.1860C}.  The MACE observations of the same activity period were also modeled with a homogeneous one-zone SSC scenario \citep{Godambe_2024ApJ...974L..31G}.  More recently, an external inverse-Compton cascade calculation was applied to the MACE and \emph{Fermi}-LAT flare data \citep{Ntshatsha_2025PoS_HEASA}.  These works show that compact leptonic or cascade scenarios are plausible.  Our goal is different: we ask whether the VLBI evidence for a disturbed parsec-scale jet can provide a physical trigger for the flare region, and we compare leptonic and pp radiation from that same reconnection site.

In this work, we test this idea with broadband SED modeling.  We describe the low-state jet emission with a stochastic-dissipation model and add a compact reconnection-powered plasmoid to represent the flaring state.  We calculate both leptonic and hadronic radiation from this flare zone.  We also use the radio constraints on the jet geometry and apparent knot speeds to estimate the Doppler factor of the TeV region.  We show that the leptonic model gives a simple explanation of the X-ray and TeV flares, while a pure pp model can also work if the flare zone contains dense gas and carries a large proton power.

The paper is organized as follows.  Section~\ref{sec:model} describes the model.  Section~\ref{sec:results} presents the SED fits and compares the leptonic and hadronic cases.  Section~\ref{sec:summary} summarizes the results.  Throughout this paper we assume $H_0=69.6\,\mathrm{km\ s^{-1}\ Mpc^{-1}}$, $\Omega_m=0.29$, and $\Omega_\Lambda=0.71$ \citep{Bennett_2014}.

\section{Model Description and Method}
\label{sec:model}

\subsection{Overall setup}

We consider two emission components in the jet.  The first component represents the quasi-steady jet emission.  It is produced by many stochastic dissipation events along the jet and accounts for the low-state broadband SED.  The second component represents the emission from the flaring zone.  It is a compact emission region produced by magnetic reconnection near the location where the parsec-scale jet changes direction. The high-state SED is contributed by the sum of the two components.
Therefore, the low-state component is mainly constrained by the archival and low-state SED.  The flare component is constrained by the X-ray, GeV, and TeV high-state data, by the TeV variability time, and by the radio measurements of the knot motion.

\subsection{Quasi-steady background emission}

We describe the low-state emission following the stochastic dissipation model developed for AGN jets \citep{Ruo_Yu_Liu_2023MNRAS.526.5054L, Wang_Ze-Rui_2022, WangZhen-Jie_2025PhRvD.112h3016W}.  In this model, the jet contains many emitting blobs.  Each blob is produced by a dissipation event at a distance $r$ from the black hole.  The ensemble of blobs forms the smooth background on which the compact TeV flare is added.

The inner jet of NGC~1275 has a viewing angle of about $20^{\circ}$--$35^{\circ}$ \citep{Park_2024A&A...685A.115P, Oh_2022MNRAS.509.1024O}, while the downstream jet may have a viewing angle of about $30^{\circ}$--$65^{\circ}$ \citep{Park_2024A&A...685A.115P, Walker_1994ApJ...430L..45W, Fujita_2017MNRAS.465L..94F}.  We therefore allow the local Doppler factor to change along the bent jet, but keep the bulk Lorentz factor fixed in the low-state calculation.  For the fiducial low-state geometry, we adopt $20^{\circ}$ for the core jet and $35^{\circ}$ for the downstream jet.  The total modeled jet length is $10\,$pc \citep{Park_2024A&A...685A.115P}.

The number of dissipation events per unit time and per unit length is parameterized as $p(r)\propto r^{-\alpha}$ \citep{Ruo_Yu_Liu_2023MNRAS.526.5054L}.  For an observing time $T$, the number of blobs in the $i$th jet segment is
\begin{equation}\label{eq:Ni}
N_i=T p(r_i)(r_{i+1}-r_i),
\end{equation}
where $r_i$ and $r_{i+1}$ are the inner and outer boundaries of the segment.  The total dissipation rate is $\dot{N}=\int_{r_0}^{r_{\rm max}}p(r)\,dr$, where $r_0$ is the jet-base distance and $r_{\rm max}$ is the outer boundary of the modeled jet.  Each blob is approximated as a sphere of radius $R(r)$ with a tangled magnetic field $B(r)$.

The Doppler factor of a blob in the $i$th segment is
\begin{equation}
\delta(r_i)=\frac{1}{\Gamma[1-\beta\cos\theta_{\rm view}(r_i)]},
\end{equation}
where $\Gamma$ is the jet bulk Lorentz factor, $\beta=(1-\Gamma^{-2})^{1/2}$, and $\theta_{\rm view}$ is the local viewing angle.  The prescriptions for $R(r_i)$, $B(r_i)$, and the blob distribution follow \citet{Ruo_Yu_Liu_2023MNRAS.526.5054L} and \citet{WangZhen-Jie_2025PhRvD.112h3016W}.  The kinetic luminosity carried by particles or magnetic field in a blob is estimated as \citep{Celotti_2008}
\begin{equation}
L_{k,i}=\pi R^2\Gamma^2\beta c U_i,
\end{equation}
where $U_i$ is the comoving energy density of electrons, protons, or magnetic field for $i=\{e,p,B\}$.

\subsection{Flare zone and jet--medium interaction}
\label{sec:flare-zone}

The TeV flare zone is motivated by the radio evidence for a rapid change in the parsec-scale jet.  VLBI observations show a knot deflection and an inverted radio spectrum that can be explained by free--free absorption \citep{Park_2024A&A...685A.115P}.  This suggests the presence of a dense, cold medium near the black hole, with an electron density above $\sim10^5\,\mathrm{cm^{-3}}$ \citep{Park_2024A&A...685A.115P}.  If the absorbing medium is clumpy, individual clouds may reach densities of $3\times10^5$--$4\times10^6\,\mathrm{cm^{-3}}$ \citep{Wajima_2020ApJ...895...35W}.  ALMA studies of other AGNs also show that the obscuring material can contain clumpy gas clouds, not only a smooth dusty disk \citep{Wada_2016ApJ...828L..19W, Izumi_2018ApJ...867...48I, Kawakatu_2020ApJ...889...84K}.

This ambient medium is important in two aspects.  First, an interaction between the jet and a cloud can compress the jet plasma and help build a current sheet, making magnetic reconnection more likely. Second, the same gas could serve as a target for pp interactions in a hadronic model. In our hadronic model, the target gas density is set as a free parameter. We do not adopt the observationally inferred gas density directly, since the inferred value is an average. The central part of the cloud could be denser and the compression by the jet could also lead to a denser target. As long as the obtained target density is not significantly (e.g., orders of magnitude) higher than the observationally inferred one, we regard it as reasonable. In general, the employed value of the target gas density is coupled with the proton luminosity, as will be discussed later.

\begin{figure}[!t]
\centering
\includegraphics[width=0.95\columnwidth]{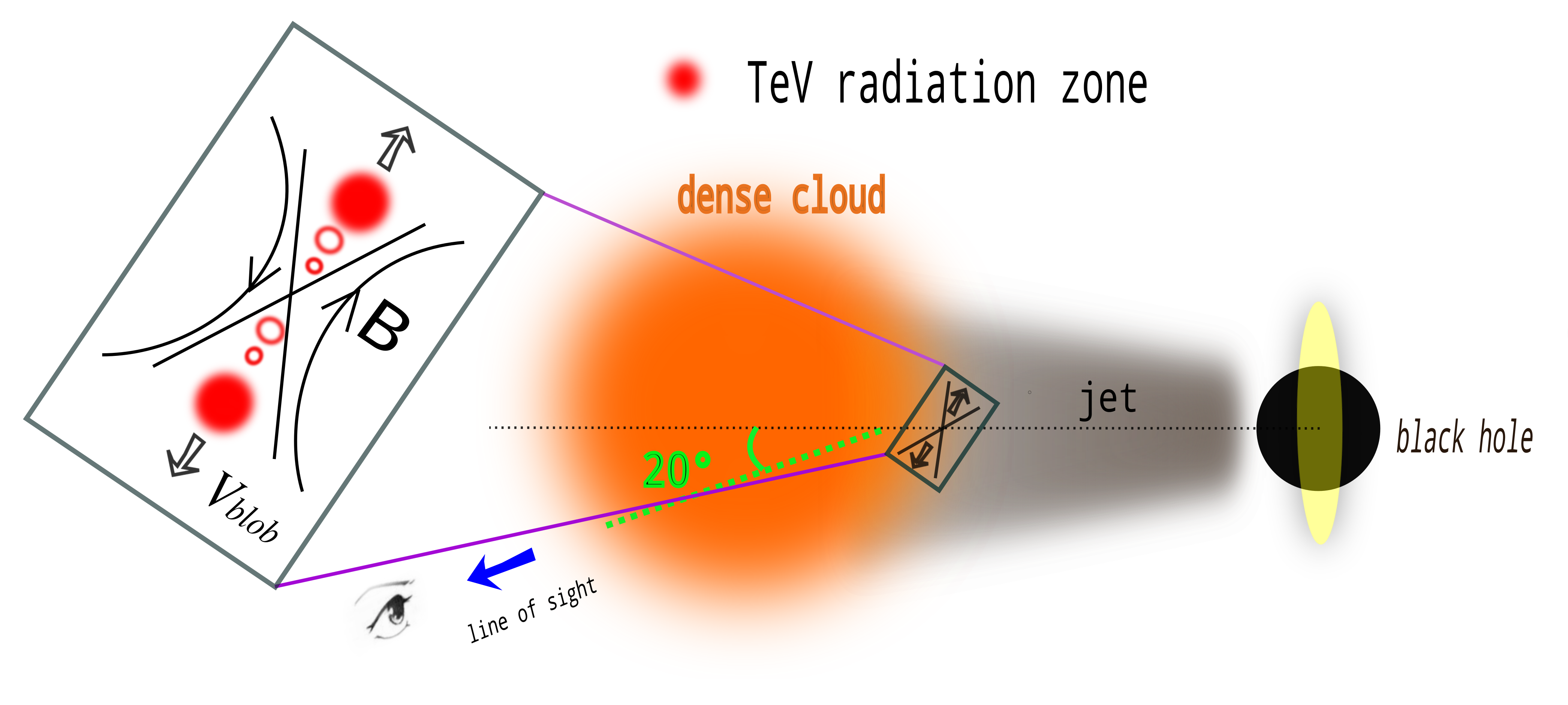}
\caption{Schematic picture of the jet--medium interaction in NGC~1275.  A dense cloud can deflect the jet and compress the magnetic field.  Magnetic reconnection in the disturbed region produces the compact TeV emission zone.  The figure is not to scale.}
\label{fig:jet_cloud}
\end{figure}

In our fiducial geometry, the jet viewing angle changes from $20^{\circ}$ in the core to $35^{\circ}$ downstream.  This gives a deflection angle of $22.4^{\circ}$ in the observer frame.  Figure~\ref{fig:jet_cloud} illustrates the physical picture, and Figure~\ref{fig:jet_geometry} shows the adopted three-dimensional geometry and its projection on the sky.  The projected geometry is chosen to be consistent with the radio image reported by \citet{Park_2024A&A...685A.115P}.

The apparent speed of the relevant knot is about $0.32c$ in the core region and $1.60c$ downstream \citep{Park_2024A&A...685A.115P}.  For a moving region with intrinsic speed $\beta c$, the apparent speed is $\beta_{\rm app}=\beta\sin\theta_{\rm view}/(1-\beta\cos\theta_{\rm view})$.  The observed apparent speeds and adopted viewing angles constrain the motion of the flare region.  As shown in Figure~\ref{fig:kinematics}, the fiducial solution gives a mildly relativistic monster plasmoid with $\Gamma_{\rm blob}\simeq1.89$ and $\delta_{\rm blob}\simeq1.7$.  The TeV zone is therefore not strongly Doppler boosted.

The downstream viewing angle is uncertain.  If it is $60^{\circ}$ instead of $35^{\circ}$, while the core viewing angle remains $20^{\circ}$, the inferred deflection angle becomes $45.3^{\circ}$ and the plasmoid Doppler factor decreases to $\delta_{\rm blob}\sim0.5$.  Across the allowed range of viewing angles, the TeV region has only weak Doppler boosting, with $\delta_{\rm blob}\sim0.5$--$1.7$.  This point is important for the hadronic model, because weak beaming implies a large intrinsic power.

\subsection{Formation of a monster plasmoid}

Magnetic reconnection can rapidly convert magnetic energy into particle energy \citep{Giannios_2009MNRAS, Giannios_2013MNRAS.431..355G, Uzdensky_2010PhRvL.105w5002U, Sironi_2025ARA&A..63..127S}.  In the plasmoid-dominated regime, many small plasmoids form in the current sheet.  A large ``monster'' plasmoid can appear if a plasmoid is born close to the center of the current sheet and stays in the reconnection layer for a long enough time.  Following \citet{Uzdensky_2010PhRvL.105w5002U}, we describe the growth of the plasmoid by its cross-sectional area $A(t)$ in the reconnection plane.  The growth rate is
\begin{equation}
\frac{dA}{dt}=y_0\exp(t/\tau_A)\widetilde{E}_{\rm eff}V_A,
\end{equation}
where $\tau_A=L/V_A$ is the global Alfv\'en crossing time, $L$ is the jet radius at the reconnection site, $y_0$ is the initial distance from the current-sheet center, $\widetilde{E}_{\rm eff}\sim0.01$ is the effective reconnection rate, and $V_A$ is the Alfv\'en speed.  We set $y_0=10^{-3}L$.

\begin{figure}[!t]
\centering
\includegraphics[width=0.95\columnwidth]{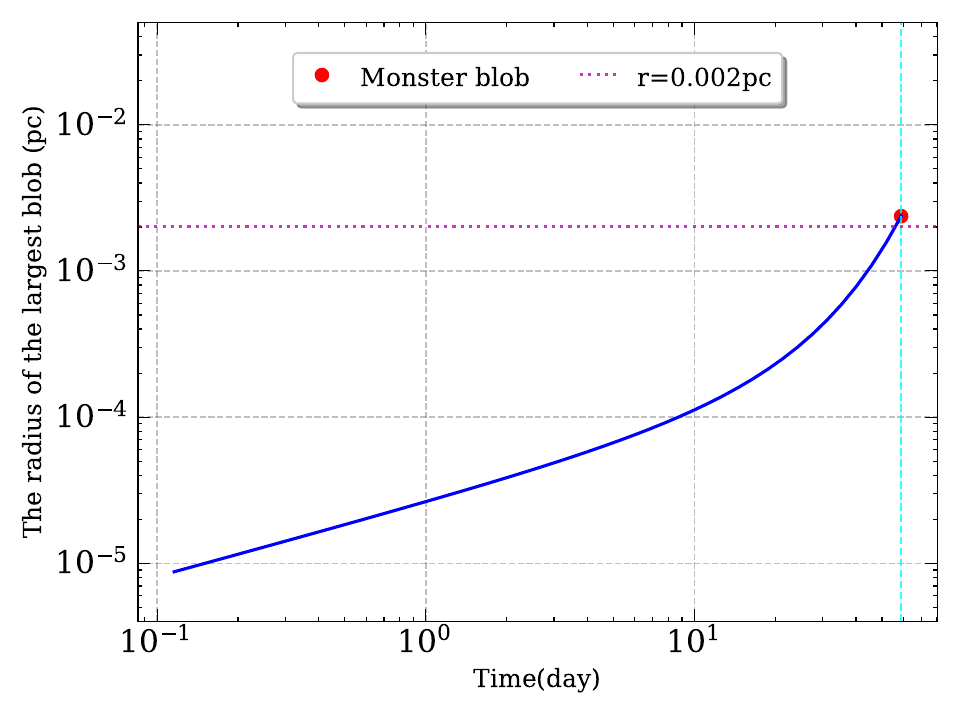}
\caption{Radius of the largest plasmoid as a function of time after the onset of reconnection.  A monster plasmoid can reach a radius of order $2\times10^{-3}\,$pc in about $60\,$days, comparable to the size inferred from the TeV variability.}
\label{fig:plasmoid_growth}
\end{figure}

The maximum plasmoid area is
\begin{equation}
A_{\rm max}=\int_0^{t_{\rm ej}}\frac{dA}{dt}\,dt,
\end{equation}
where $t_{\rm ej}\sim\tau_A\ln(L/y_0)$ is the ejection time for a plasmoid born at $y_0$ \citep{Uzdensky_2010PhRvL.105w5002U}.  We estimate the characteristic plasmoid radius as $R_{\rm blob}=(A_{\rm max}/\pi)^{1/2}$.  Figure~\ref{fig:plasmoid_growth} shows that a monster plasmoid can grow to a radius of order $2\times10^{-3}\,$pc in about $60\,$days.  This timescale is close to the interval between the radio knot acceleration and the TeV outbursts.

The bulk Lorentz factor of the monster plasmoid in the lab frame is \citep{Giannios_2009MNRAS}
\begin{equation}
\Gamma_{\rm blob}=\Gamma\Gamma'_{\rm blob}(1+\beta\beta'_{\rm blob}\cos\theta'),
\end{equation}
where $\beta c$ is the velocity of the underlying jet, and $\Gamma'_{\rm blob}$, $\beta'_{\rm blob}c$, and $\theta'$ are the Lorentz factor, velocity, and propagation angle of the plasmoid measured in the jet frame.  The corresponding deflection angle in the lab frame is
\begin{equation}
\tan\theta=\frac{\beta'_{\rm blob}\sin\theta'}{\Gamma(\beta'_{\rm blob}\cos\theta'+\beta)}.
\end{equation}

The observed TeV flare duration of several days gives a variability size
\begin{equation}
R_{\rm var}\lesssim \frac{c\Delta t\delta_{\rm blob}}{1+z}.
\end{equation}
For the fiducial $\delta_{\rm blob}\simeq1.7$, this gives $R\simeq0.001\,$pc for O1 and $R\simeq0.003\,$pc for O2. These sizes are consistent with the monster-plasmoid size estimated above. 

\subsection{Particle evolution and radiation}

Electrons and, in the hadronic case, protons are injected into each emission zone at a constant rate for a time $t_{\rm inj}=R(r_i)/c$.  The injection spectrum is a power-law or smoothly broken power law,
\begin{equation}
\begin{aligned}
Q_{\rm e/p}(\gamma_{\rm e/p})=&Q_{\rm e/p,0}\gamma_{\rm e/p}^{-n_{\rm e/p}},\\
Q_{\rm e/p}(\gamma_{\rm e/p})=&Q_{\rm e/p,0}\gamma_{\rm e/p}^{-n_{\rm e/p,1}}\\
&\times\left[1+\left(\frac{\gamma_{\rm e/p}}{\gamma_{\rm e/p,b}}\right)^{n_{\rm e/p,2}-n_{\rm e/p,1}}\right]^{-1}.
\end{aligned}
\end{equation}
The normalization is set by $\int Q_{\rm e/p}\gamma_{\rm e/p}m_{\rm e/p}c^2\,d\gamma_{\rm e/p}=L_{\rm e/p,inj}/[(4/3)\pi R^3]$.  The particle distribution $N_{\rm e/p}(\gamma_{\rm e/p},t,r_i)$ evolves as
\begin{equation}
\begin{aligned}
\frac{\partial N_{\rm e/p}(\gamma_{\rm e/p},t,r_i)}{\partial t}=&-\frac{\partial}{\partial \gamma}\left[\dot{\gamma}(\gamma,t,r_i)N_{\rm e/p}(\gamma_{\rm e/p},t,r_i)\right]\\
&-\frac{N_{\rm e/p}(\gamma_{\rm e/p},t,r_i)}{t_{\rm esc}}+Q_{\rm e/p}(\gamma_{\rm e/p}),
\end{aligned}
\end{equation}
where $t_{\rm esc}=10R(r_i)/c$ \citep{Gao_2017ApJ...843..109G}.  The total energy-loss rate is
\begin{equation}
\dot{\gamma}(\gamma,t,r_i)=-\frac{\gamma}{t_{\rm dyn}}-\frac{\gamma}{t_{\rm rad}},
\end{equation}
with $t_{\rm dyn}=R(r_i)/c$.

For leptonic emission, the radiative cooling time is
\begin{equation}
t_{\rm rad}=\frac{3m_{\rm e}c}{4(U_B+\kappa_{\rm KN}U_{\rm ph})\sigma_T\gamma_{\rm e}},
\end{equation}
where $U_B=B^2/8\pi$, $U_{\rm ph}$ is the soft-photon energy density, and $\kappa_{\rm KN}$ accounts approximately for Klein--Nishina suppression \citep{2005MNRAS.363..954M}.  The soft-photon field includes synchrotron photons from the same blob, photons from neighboring blobs, and external photons from the broad-line region (BLR), and dusty torus (DT).  A cooling break appears when $t_{\rm rad}(\gamma)=t_{\rm dyn}$.

For hadronic emission, the proton cooling time is approximated as $t_{\rm rad}=\min\{t_{p\gamma},t_{\rm BH},t_{pp},t_{p,{\rm syn}}\}$.  The maximum proton energy is obtained from
\begin{equation}
t_{\rm acc}=\min\{t_{\rm dyn},t_{p\gamma},t_{\rm BH},t_{pp},t_{p,{\rm syn}}\},
\end{equation}
where the acceleration time is
\begin{equation}
t_{\rm acc}\simeq\frac{\alpha r_L}{c}\simeq\frac{\alpha\gamma_p m_p c}{eB}.
\end{equation}
We take $\alpha=10$, noting that $\alpha=1$ corresponds to Bohm-limit acceleration \citep{Rieger_2007Ap&SS}.

For the external radiation field, we include BLR and DT photons and approximate both as isotropic blackbody components in the AGN frame.  Their characteristic frequencies in the jet comoving frame peak near $2\times10^{15}\Gamma\,\mathrm{Hz}$ for the BLR \citep{2008MNRAS.386..945T} and $3\times10^{13}\Gamma\,\mathrm{Hz}$ for the DT \citep{2007ApJ...660..117C}.  The energy densities are calculated as \citep{2012ApJ...754..114H}
\begin{equation}
u_{\rm BLR}=\frac{\eta_{\rm BLR}\Gamma^2L_d}{3\pi r_{\rm BLR}^2c[1+(r/r_{\rm BLR})^3]},
\end{equation}
\begin{equation}
u_{\rm DT}=\frac{\eta_{\rm DT}\Gamma^2L_d}{3\pi r_{\rm DT}^2c[1+(r/r_{\rm DT})^4]},
\end{equation}
where $\eta_{\rm BLR}=\eta_{\rm DT}=0.1$.  We use $r_{\rm BLR}=0.1(L_d/10^{46}\,\mathrm{erg\ s^{-1}})^{1/2}\,\mathrm{pc}$ and $r_{\rm DT}=2.5(L_d/10^{46}\,\mathrm{erg\ s^{-1}})^{1/2}\,\mathrm{pc}$.  For NGC~1275, these values give $r_{\rm BLR}=0.006\,$pc and $r_{\rm DT}=0.15\,$pc.  The BLR luminosity is taken to be $0.1L_d=3.8\times10^{42}\,\mathrm{erg\ s^{-1}}$.  The photopion, Bethe--Heitler, and cascade calculations follow \citet{2008PhRvD..78c4013K} and \citet{Bottcher_2013}.

Internal $\gamma\gamma$ absorption is included through the optical depth
\begin{equation}
\tau_{\gamma\gamma}(\varepsilon_1)=\frac{R\pi r_e^2}{\varepsilon_1^2}\int_{1/\varepsilon_1}^{\infty}d\varepsilon\,n_{\rm soft}(\varepsilon)\bar{\phi}(s_0)\varepsilon^{-2},
\end{equation}
where $\varepsilon_1$ and $\varepsilon$ are the dimensionless energies of high and low energy photons, $s_0=\varepsilon\varepsilon_1$. $\bar{\phi}(s_0)$ can be given by \cite{Xue_2022PhRvD.106j3021X}. $R$ is the size of the flare zone, $r_e$ is the classical electron radius, and $n_{\rm soft}$ is the total soft-photon density. This density includes photons produced inside the flare zone and external photons from the BLR and DT. In the fiducial fits, the internal photon field of the TeV region dominates the absorption. 

\FloatBarrier

\section{SED Modeling and Results}
\label{sec:results}
The Equation (10) shows the temporal evolution of the electron (proton) spectrum in one blob. The injection time $t_{\rm inj}$ of particles in each blob is set equal to the light crossing time $R(r_{i})/c$, where $r_{i}$ is the distance of the $i$th segment to the black hole. After the cease of particle injection, particles confined within a blob will lose energy due to adiabatic expansion after several light crossing times, even if the radiative cooling is not important. So for each blob, we solve the equation up to $t=10R(r_{i})/c$. The low-state (or steady-state) SED of the $i$th segment is calculated by summing the emission of $N_i$ blobs (given by Eq.~\ref{eq:Ni}) generated in this segment over the observation period $T$ and dividing it by $T$ (more details can be found in Ref.~\cite{Ruo_Yu_Liu_2023MNRAS.526.5054L}). The total steady-state SED of the entire jet is the sum of the steady-state emission of all segments. 

We first fit the low-state broadband emission and then add the compact flare component.  The low-state parameters do not set the intrinsic properties of the TeV plasmoid, but they determine the background level on which the flare is superposed.  The jet base is placed at $r_0\simeq0.06\,\mathrm{pc}$, corresponding to $\sim560r_{\rm Sch}$, where $r_{\rm Sch}$ is the Schwarzschild radius.  Since the corona is usually expected to extend only to $\sim30r_{\rm Sch}$, coronal photons do not efficiently interact with the electrons or protons at the jet base in our model.

For the low-state jet component, we adopt $B_0\simeq1\,\mathrm{G}$ at the jet base and a bulk Lorentz factor $\Gamma\simeq2.5$.  The total blob generation rate is $\dot{N}=0.4\,\mathrm{s^{-1}}$, the dissipation-probability index is $\alpha=0.5$, and the ratio of the blob radius to the segment radius is $\kappa=0.01$.  The injected electron spectrum has $\gamma_{\rm e,min}=1$, $\gamma_{\rm e,break}=7\times10^4$, $\gamma_{\rm e,max}=10^7$, $n_{\rm e,1}=2.5$, $n_{\rm e,2}=3.5$, and $L_{\rm e,inj}=4.5\times10^{42}\,\mathrm{erg\ s^{-1}}$.  The peak electron kinetic luminosity and magnetic-field kinetic luminosity along the jet are $1.0\times10^{46}\,\mathrm{erg\ s^{-1}}$ and $3.8\times10^{42}\,\mathrm{erg\ s^{-1}}$, respectively.

Table~\ref{tab:parameters} summarizes the parameters of the compact TeV region.  We list O1 and O2 separately because the two outbursts have different variability sizes and slightly different luminosities, while the basic geometry is kept the same.

\begin{table*}[!t]
\centering
\caption{Summary of parameters for the compact TeV emission zone.}
\label{tab:parameters}

\begingroup
\renewcommand{\arraystretch}{1.08}
\setlength{\tabcolsep}{8pt}
\begin{tabular}{l c c l}
\toprule
Parameter & ~~~O1~~~ & O2 & Description \\
\midrule
\multicolumn{4}{c}{TeV emission zone} \\
\midrule
$R\ (\mathrm{pc})$ & $0.001$ & $0.003$ & Radius of the flare zone \\
$\delta$ & $1.7$ & $1.7$ & Doppler factor of the flare zone \\
$\Gamma$ & $1.89$ & $1.89$ & Bulk Lorentz factor of the flare zone \\
\midrule
\multicolumn{4}{c}{Leptonic scenario} \\
\midrule
$B\ (\mathrm{G})$ & $0.1$ & $0.1$ & Magnetic field in the flare zone \\
$n_{\rm e}$ & $2.0$ & $2.0$ & Electron spectral index \\
$\gamma_{\rm e,min}$ & $1.0\times10^5$ & $1.0\times10^5$ & Minimum electron Lorentz factor \\
$\gamma_{\rm e,max}$ & $6\times10^7$ & $6\times10^7$ & Maximum electron Lorentz factor \\
$L_{\rm e,inj}\ (\mathrm{erg\ s^{-1}})$ & $6.5\times10^{44}$ & $6.5\times10^{44}$ & Electron injection luminosity \\
$L_{\rm k,e}\ (\mathrm{erg\ s^{-1}})$ & $4.8\times10^{43}$ & $2.9\times10^{43}$ & Electron kinetic luminosity \\
$L_{\rm k,B}\ (\mathrm{erg\ s^{-1}})$ & $1.1\times10^{39}$ & $9.7\times10^{39}$ & Magnetic kinetic luminosity \\
\midrule
\multicolumn{4}{c}{Hadronic scenario} \\
\midrule
$B\ (\mathrm{G})$ & $8.0$ & $8.0$ & Magnetic field in the flare zone \\
$n_{\rm H}\ (\mathrm{cm^{-3}})$ & $2.0\times10^7$ & $2.0\times10^7$ & Cold target density for pp interactions \\
$n_{\rm p,1}$ & $1.6$ & $2.0$ & Low-energy proton spectral index \\
$n_{\rm p,2}$ & $3.8$ & $3.8$ & High-energy proton spectral index \\
$\gamma_{\rm p,min}$ & $1.0$ & $1.0$ & Minimum proton Lorentz factor \\
$\gamma_{\rm p,break}$ & $2.0\times10^4$ & $3.0\times10^4$ & Break proton Lorentz factor \\
$\gamma_{\rm p,max}$ & $7.8\times10^8$ & $2.3\times10^9$ & Maximum proton Lorentz factor \\
$L_{\rm p,inj}\ (\mathrm{erg\ s^{-1}})$ & $8.0\times10^{47}$ & $5.0\times10^{47}$ & Proton injection luminosity \\
$n_{\rm e,1}$ & $2.0$ & $2.0$ & Low-energy electron spectral index \\
$n_{\rm e,2}$ & $3.5$ & $3.5$ & High-energy electron spectral index \\
$\gamma_{\rm e,min}$ & $1\times10^4$ & $1\times10^4$ & Minimum electron Lorentz factor \\
$\gamma_{\rm e,break}$ & $5\times10^6$ & $5\times10^6$ & Break electron Lorentz factor \\
$\gamma_{\rm e,max}$ & $1.3\times10^7$ & $1.3\times10^7$ & Maximum electron Lorentz factor \\
$L_{\rm e,inj}\ (\mathrm{erg\ s^{-1}})$ & $1.0\times10^{44}$ & $1.0\times10^{44}$ & Electron injection luminosity \\
$L_{\rm k,e}\ (\mathrm{erg\ s^{-1}})$ & $7.01\times10^{40}$ & $2.61\times10^{40}$ & Electron kinetic luminosity \\
$L_{\rm k,p}\ (\mathrm{erg\ s^{-1}})$ & $3.07\times10^{47}$ & $1.76\times10^{47}$ & Proton kinetic luminosity \\
$L_{\rm k,B}\ (\mathrm{erg\ s^{-1}})$ & $6.92\times10^{42}$ & $6.23\times10^{43}$ & Magnetic kinetic luminosity \\
\bottomrule
\end{tabular}
\endgroup
\end{table*}

In the low state, the stochastic dissipation component reproduces the multi-wavelength jet emission.  Its inverse-Compton emission includes both synchrotron self-Compton (SSC) and external Compton (EC) scattering, but SSC dominates.  The host-galaxy component accounts for the optical and infrared excess.  We use a Seyfert-galaxy SWIRE template for the host contribution.

\begin{figure*}[!t]
\centering
\begin{minipage}{0.49\textwidth}
\centering
\includegraphics[width=0.98\linewidth]{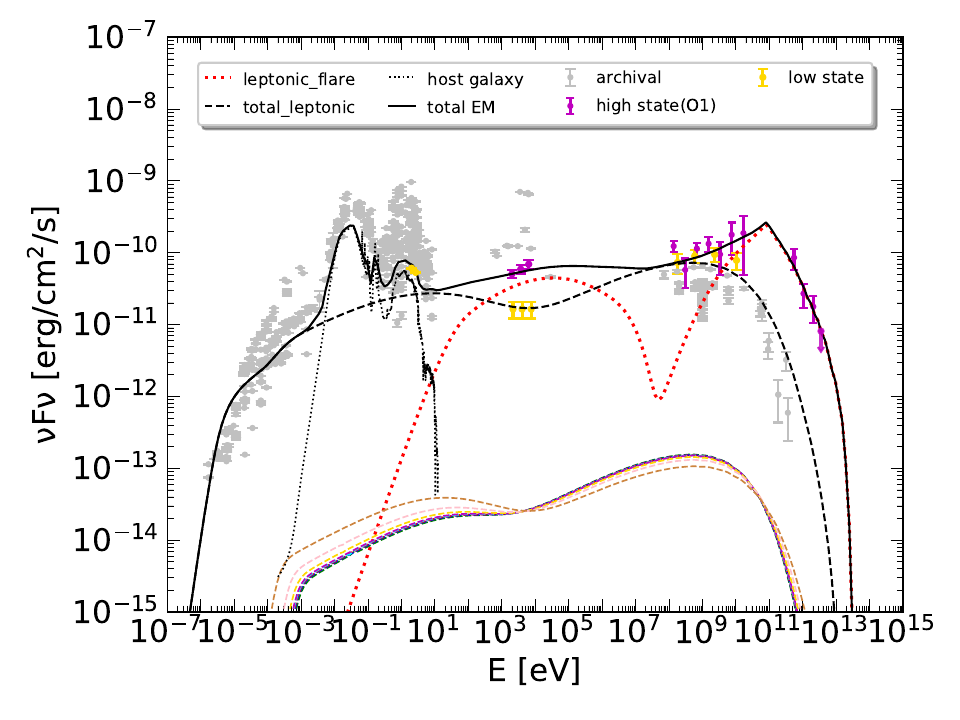}
\end{minipage}
\hfill
\begin{minipage}{0.49\textwidth}
\centering
\includegraphics[width=0.98\linewidth]{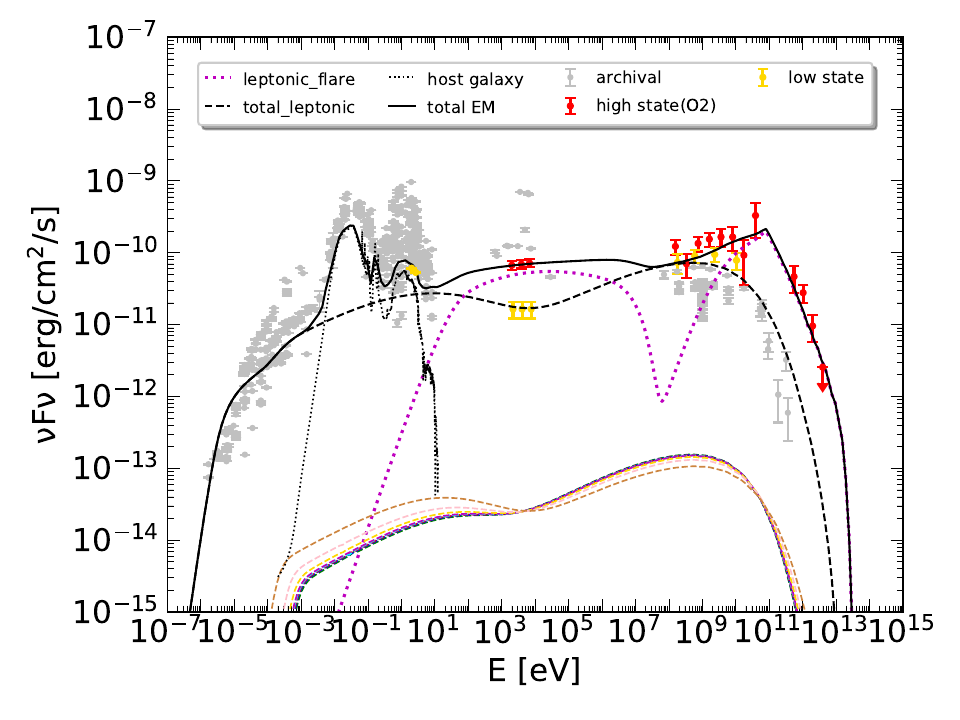}
\end{minipage}
\caption{Broadband SED modeling in the leptonic scenario.  Left: O1.  Right: O2.  Gray points are historical archival data from the ASI/SSDC SED Builder.  Optical data are from the Zwicky Transient Facility.  The low- and high-state X-ray data and the low-state $\gamma$-ray data are from \citet{Godambe_2024ApJ...974L..31G}; the high-state GeV and TeV data are from \citet{Caozhen_2025MNRAS.540.1860C}.  The red/magenta dotted curve is the emission from the reconnection-powered monster plasmoid.}
\label{fig:sed_leptonic}
\end{figure*}

The leptonic flare fits are shown in Figure~\ref{fig:sed_leptonic}.  A compact monster plasmoid can reproduce the high-state X-ray and TeV emission.  In this region, the magnetic field is $B\simeq0.1\,\mathrm{G}$.  The injected electron spectrum has $n_{\rm e}\simeq2.0$, $\gamma_{\rm e,min}=10^5$, and $\gamma_{\rm e,max}=6\times10^7$.  The X-ray flare is dominated by synchrotron radiation from these high-energy electrons, while the TeV emission is mainly produced by IC scattering.  The same parameter set works for both O1 and O2 after changing the size and luminosity of the flare region.  This result is broadly consistent with earlier SSC modeling of the same activity period \citep{Caozhen_2025MNRAS.540.1860C, Godambe_2024ApJ...974L..31G}, but our model ties the compact emission zone to a possible reconnection event in the bent parsec-scale jet.

\begin{figure*}[!t]
\centering
\begin{minipage}{0.49\textwidth}
\centering
\includegraphics[width=0.98\linewidth]{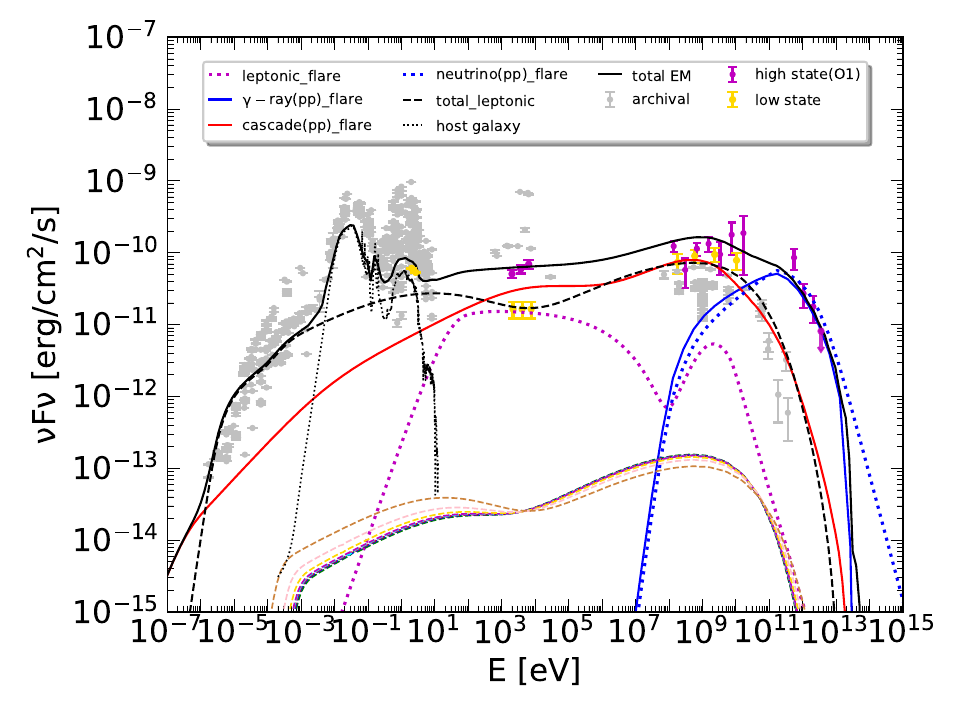}
\end{minipage}
\hfill
\begin{minipage}{0.49\textwidth}
\centering
\includegraphics[width=0.98\linewidth]{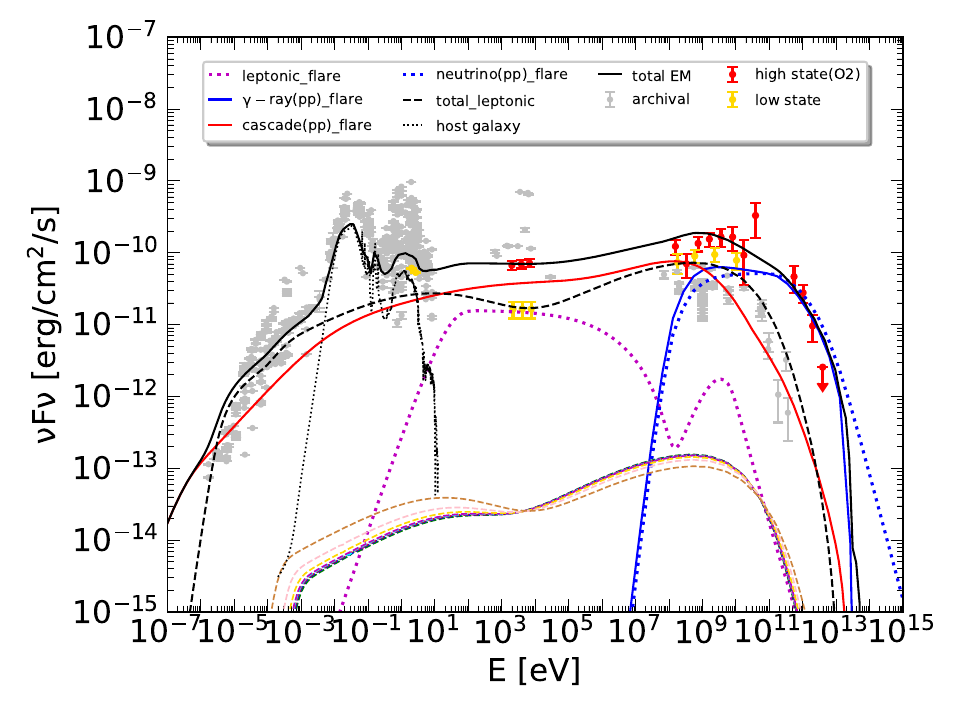}
\end{minipage}
\caption{Broadband SED modeling in the hadronic scenario.  Left: O1.  Right: O2.  Dashed curves show the leptonic background emission.  The blue solid curve shows $\gamma$ rays from $\pi^0$ decay in pp interactions, the blue dotted curve shows the accompanying $\nu_\mu+\bar{\nu}_\mu$ emission, the red solid curve shows electromagnetic cascade emission, and the magenta dotted curve shows synchrotron radiation from primary electrons in the monster plasmoid.}
\label{fig:sed_hadronic}
\end{figure*}

\begin{figure}[htbp]
\centering
\includegraphics[width=0.95\columnwidth]{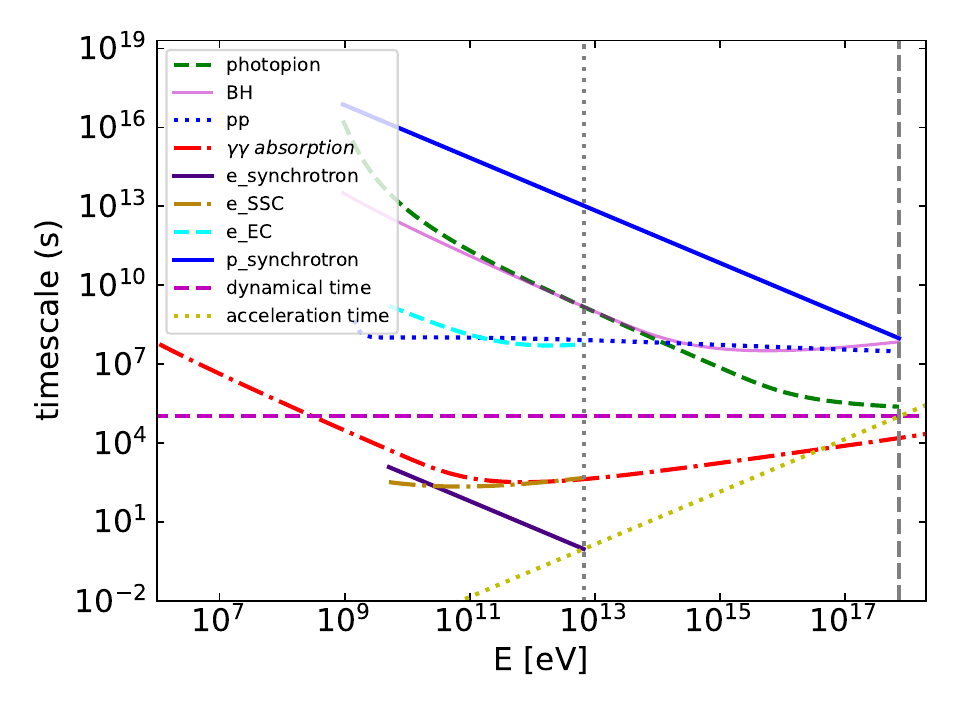}
\caption{Cooling, acceleration, and absorption timescales in the comoving frame of the monster plasmoid for the hadronic scenario.  The radius is set to $\sim0.001\,$pc.  The vertical dashed line marks the maximum proton energy set by $t_{\rm acc}=\min\{t_{\rm cool},t_{\rm p,dyn}\}$, and the dotted line marks the maximum electron energy adopted in the model.}
\label{fig:hadronic_timescales}
\end{figure}

The hadronic fits are shown in Figure~\ref{fig:sed_hadronic}.  In this case, TeV photons are produced mainly by pp interactions and by the cascade emission initiated by secondary particles.  Figure~\ref{fig:hadronic_timescales} shows the relevant cooling and absorption timescales in the monster plasmoid.  Protons below $\sim100\,\mathrm{TeV}$ cool mainly through pp interactions, while photopion production becomes faster at higher energies.  However, the proton spectrum breaks at a few times $10^4$ in Lorentz factor and has a steep high-energy index, so the photopion contribution to the observed SED remains weak.

The pp model needs dense cold gas in the flare zone.  The target density used in the SED fit is $n_{\rm H}\simeq2\times10^7\,\mathrm{cm^{-3}}$.  This is above the $3\times10^5$--$4\times10^6\,\mathrm{cm^{-3}}$ range inferred for clumpy free--free absorbing clouds by \citet{Wajima_2020ApJ...895...35W}, but the difference is not decisive.  The observed absorber may trace an average column rather than the densest part of a cloud.  In addition, a cloud core can be denser than the value inferred from free--free absorption, and a jet--cloud interaction can further compress the gas.

The other demanding parameter is the proton kinetic luminosity, $L_{k,p}\sim10^{47}\,\mathrm{erg\ s^{-1}}$.  This is comparable to the Eddington luminosity of NGC~1275,
\begin{equation}
L_{\rm Edd}\simeq1.26\times10^{47}\left(\frac{M_{\rm BH}}{10^9M_\odot}\right)\,\mathrm{erg\ s^{-1}}.
\end{equation}
However, the target density and proton luminosity are coupled.  For a fixed TeV luminosity, a higher $n_{\rm H}$ increases the pp efficiency and reduces the required proton power, while a lower $n_{\rm H}$ does the opposite.  The Eddington luminosity is also not a strict upper limit for the transient power of a relativistic jet during a flare.  We therefore do not rule out the pp interpretation, but regard it as a possible scenario with demanding parameters.

The adopted magnetic field at the jet base, $B_0\simeq1\,\mathrm{G}$, is close to the $1.8$--$4.0\,\mathrm{G}$ field strength measured at the jet apex by \citet{Paraschos_2021A&A...650L..18P}, although it is below the lower limit of $21\pm14\,\mathrm{G}$ reported by \citet{Kim_2019A&A...622A.196K}.  Given the uncertainties in the magnetic-field estimates and in the jet geometry, the low-state SED can be reproduced with the adopted parameters.  The leptonic model requires less demanding matter density and proton power, while the pp model remains viable if the local gas conditions are more extreme than those directly inferred from the radio absorber.

%\FloatBarrier

\section{Summary}
\label{sec:summary}

We have studied the origin of the TeV $\gamma$-ray outbursts detected from NGC~1275 by LHAASO between November 2022 and January 2023.  The model is motivated by the radio evidence for a sudden knot acceleration and jet deflection in late 2022 \citep{Park_2024A&A...685A.115P}.  We considered a scenario in which the jet interacts with dense ambient gas.  This interaction can compress the magnetic field, trigger magnetic reconnection, and produce a compact monster plasmoid.

The radio knot speeds and the allowed viewing angles constrain the motion of the TeV flare zone.  For the fiducial geometry with a $20^{\circ}$ core viewing angle and a $35^{\circ}$ downstream viewing angle, we obtain $\Gamma_{\rm blob}\simeq1.89$ and $\delta_{\rm blob}\simeq1.7$.  The observed flare durations imply radii of $R\simeq0.001\,$pc for O1 and $R\simeq0.003\,$pc for O2.  These sizes are consistent with a monster plasmoid that grows over about two months, similar to the interval between the radio knot acceleration and the TeV outbursts.

We modeled the broadband SEDs with both leptonic and hadronic flare components.  In the leptonic model, the high-state X-ray emission is synchrotron radiation from freshly accelerated electrons, and the TeV emission is mainly IC radiation from the same compact region.  A magnetic field of $B\simeq0.1\,\mathrm{G}$ is enough to reproduce the X-ray and TeV data.

The pure pp model can also reproduce the TeV component, but it is more demanding.  It requires a cold target density of $\sim2\times10^7\,\mathrm{cm^{-3}}$ and a proton kinetic luminosity of order $10^{47}\,\mathrm{erg\ s^{-1}}$.  The required density is above the value directly inferred from radio free--free absorption, and the required proton power is close to the Eddington luminosity of the central black hole.  These facts should be interpreted as pressure on the model parameters, not as a strict exclusion.  Dense cloud cores, gas compression during the jet--cloud interaction, or a temporarily super-Eddington jet power during the flare could make the pp scenario possible.

We therefore conclude that magnetic reconnection in the parsec-scale jet is a viable origin of the 2022--2023 TeV outbursts.  The leptonic interpretation is simpler in terms of energetics, while the hadronic interpretation remains an interesting but more extreme possibility.  Future simultaneous VLBI, X-ray, GeV, and TeV monitoring will be important for testing this picture.  A key test is whether future knot accelerations or jet deflections are followed by high-energy flares on a similar timescale.  Better constraints on the density, size, and covering factor of the ambient gas will also be crucial for testing the pp interpretation.

\section*{Acknowledgments}

This work is supported by National Science Foundation of China under grants No.~12393852 and 12333006, and Basic Research Program of Jiangsu under grant No.~BK20250059.

\vspace{5mm}
\bibliographystyle{apsrev}
\bibliography{ms}

\appendix
\section*{Appendix}
\setcounter{figure}{0}
\renewcommand{\thefigure}{A\arabic{figure}} 

\begin{figure*}[!t]
\centering
\begin{minipage}{0.49\textwidth}
\centering
\includegraphics[width=0.98\linewidth]{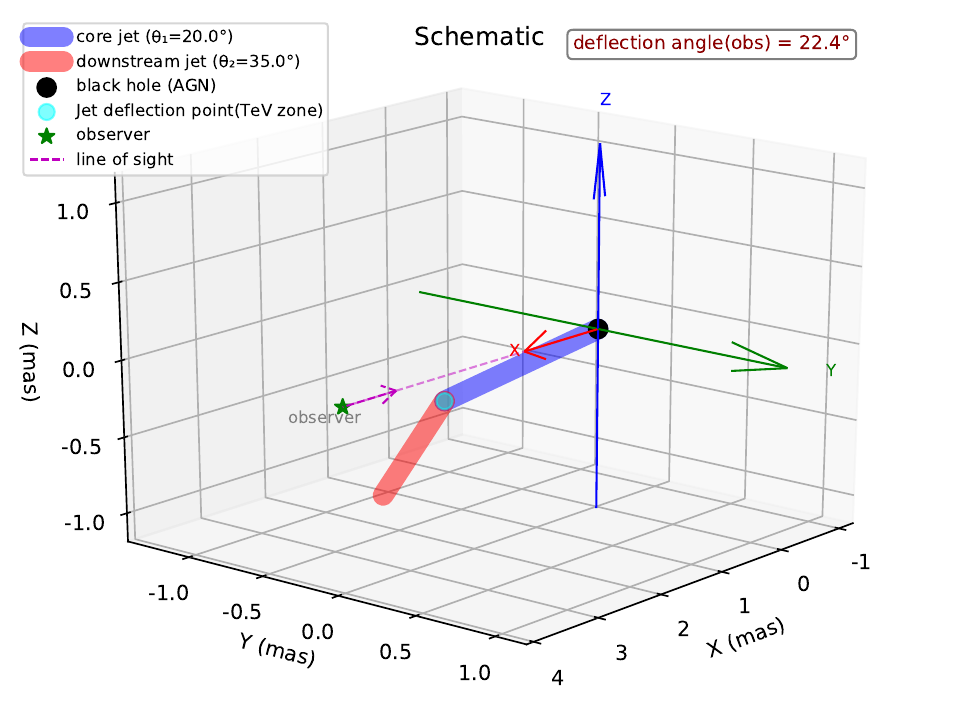}
\end{minipage}
\hfill
\begin{minipage}{0.49\textwidth}
\centering
\includegraphics[width=0.98\linewidth]{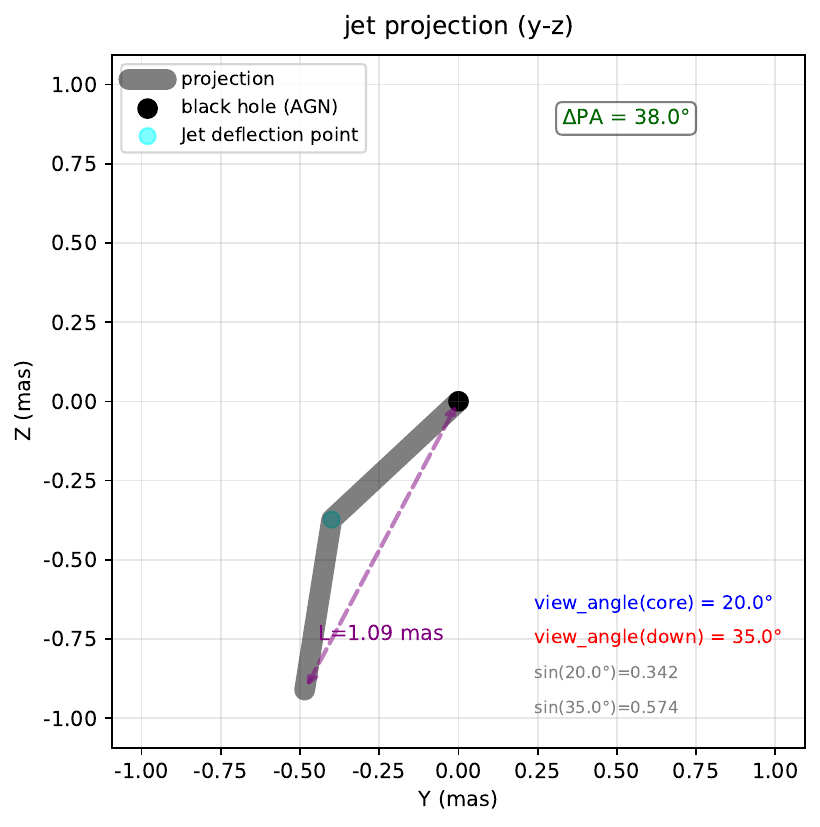}
\end{minipage}
\caption{Geometry of the bent jet used in the calculation.  Left: three-dimensional schematic in the observer frame.  The blue component is the core jet, the red component is the downstream jet, the cyan point marks the jet-deflection and TeV-emission region, and the green star marks the observer.  Right: projection on the sky plane.  The projected position angle change is consistent with the radio morphology of NGC~1275.}
\label{fig:jet_geometry}
\end{figure*}

\begin{figure*}[!t]
\centering
\begin{minipage}{0.49\textwidth}
\centering
\includegraphics[width=0.98\linewidth]{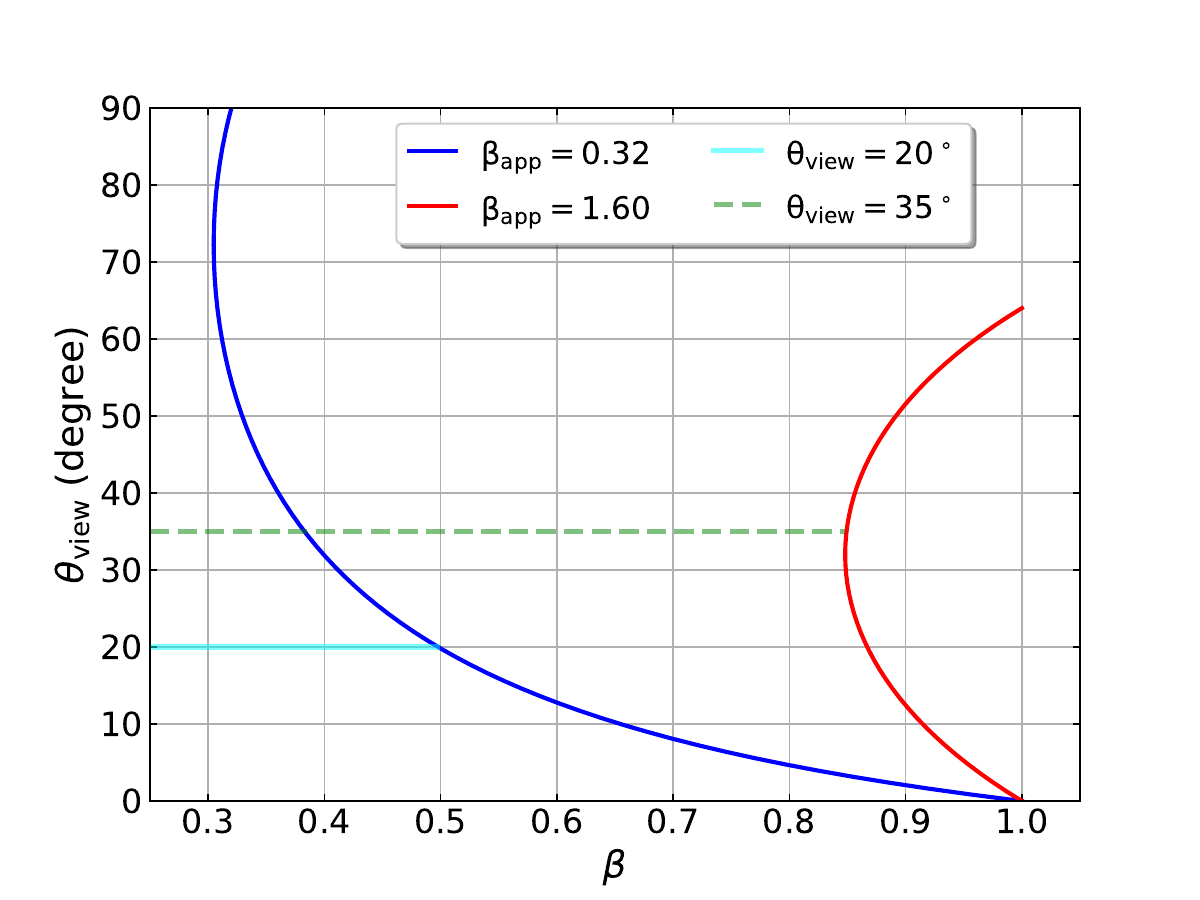}
\end{minipage}
\hfill
\begin{minipage}{0.49\textwidth}
\centering
\includegraphics[width=0.98\linewidth]{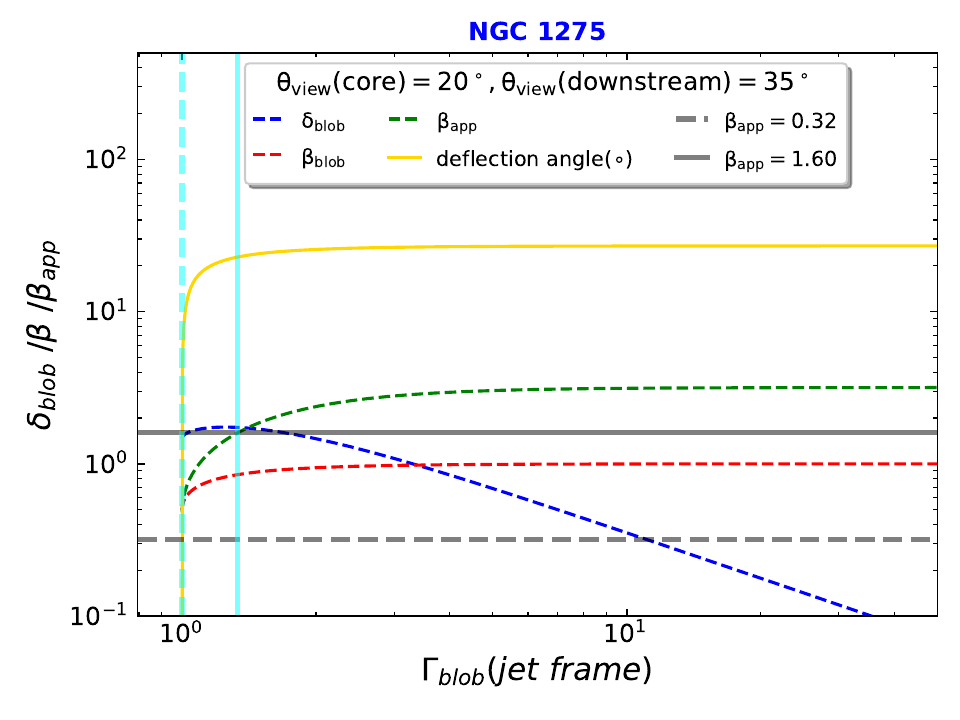}
\end{minipage}
\caption{Kinematic constraints on the compact TeV emission zone.  Left: apparent speed as a function of intrinsic speed and viewing angle.  The blue and red curves mark $\beta_{\rm app}=0.32$ and $\beta_{\rm app}=1.60$, respectively.  Right: Doppler factor, speed, and apparent speed of the monster plasmoid.  The fiducial geometry gives $\Gamma_{\rm blob}\simeq1.89$ and $\delta_{\rm blob}\simeq1.7$.}
\label{fig:kinematics}
\end{figure*}

Here we present two figures that support the geometrical and kinematic arguments in the main text. Figure~\ref{fig:jet_geometry} shows the adopted three‑dimensional geometry and its projection onto the sky. Figure~\ref{fig:kinematics} illustrates the kinematic constraints on the compact TeV emission zone.

\end{document}